\documentstyle[prb,aps,twocolumn]{revtex}
\begin{document}

\title{Unoccupied Band Structure of NbSe$_2$ by 
Very-Low-Energy Electron Diffraction: 
Experiment and Theory
}
\author{E. E. Krasovskii and W.~Schattke}
\address{Institut f\"ur Theoretische Physik,
Christian-Albrechts-Universit\"at,\\  
Leibnizstrasse 15, D-24098 Kiel, Germany}
\author{V. N. Strocov \cite{IHPC} and R. Claessen}
\address{Experimentalphysik II, Universit\"at Augsburg, 
D-86135 Augsburg, Germany}

\maketitle 

\begin{abstract} 
A combined experimental and theoretical study of very-low-energy
electron diffraction at the (0001) surface of 2H-NbSe$_2$ is
presented. Electron transmission spectra have been measured for
energies up to 50~eV above the Fermi level with ${\bf k}^{||}$
varying along the $\Gamma$K line of the Brillouin zone. 
{\it Ab initio} calculations of the spectra have been performed with 
the extended linear augmented plane wave $\mathbf{k}\cdot\mathbf{p}$
method. The experimental spectra are interpreted in terms of 
three-dimensional ${\bf k}^{||}$-resolved one-electron band 
structure. Special attention is paid to the quasi-particle lifetimes:
by comparing the broadening of the spectral structures in the
experimental and calculated spectra the energy dependence of the 
optical potential $-iV_{\rm i}$ is determined. A sharp increase of 
$V_{\rm i}$ at 20~eV is detected, which is associated with a plasmon 
peak in the Im$[-1/\varepsilon]$ function. Furthermore, the electron 
energy loss spectrum and the reflectivity of NbSe$_2$ are calculated 
{\it ab initio} and compared with optical experiments. The obtained 
information on the dispersions and lifetimes of the unoccupied states 
is important for photoemission studies of the 3D band structure of 
the valence band.
\end{abstract} 

\section{introduction} 
Very-Low-Energy Electron Diffraction (VLEED) spectroscopy with
incident electron energies below ${\sim 40}$~eV has recently been
established as the experimental method giving a direct access to
dispersions $E({\bf k})$ and lifetimes of quasi-particle states
above the vacuum level 
(see Refs.~\onlinecite{StrocovExi,StrocovTMDCs,StrocovCFS,Peetz92} and
references therein). The VLEED spectral structures reveal the critical
points in the surface-perpendicular dispersions $E(k^{\perp})$ of the
states that couple to the incident electron beam to form the LEED
state. The fundamental advantage of the VLEED spectroscopy over the
conventional unoccupied band structure methods, such as inverse
photoemission or X-ray absorption spectroscopy, is that it does not
involve initial states and in particular cases, for example at
high-symmetry lines parallel to the surface, is capable of restoring
3D Bloch vectors ${\bf k}$.

Within a quasi-particle picture \cite{Feibelman74,Kevan92} the 
time-reversed LEED states serve as the final states of the one-step 
photoemission theory. Independent information on their dispersions and 
lifetimes provided by the VLEED experiment can be used in the 
photoemission experiment to resolve the valence band structure in the 
surface-perpendicular direction, \cite{StrocovTMDCs,StrocovCFS} which is 
often blurred by the complicated non-free-electron-like dispersion and 
strong self-energy effects in the final states. In particular, the 
broadening of the VLEED spectral structures provides information on the 
lifetimes of the final state quasi-particles, which can be used to estimate 
the $k^{\perp}$ broadening of the Bloch constituents of the LEED state and 
thereby to judge on the intrinsic uncertainty of the band mapping.

Within a simplified one-particle approach, the LEED problem reduces to
solving the Schr\"odinger equation for a semi-infinite crystal given
the energy $E$ and the initial conditions of the incident electron, 
i.e., the asymptotics of the wave function in the vacuum. In the plane 
parallel to the crystal surface the LEED wave function obeys the Bloch 
theorem and is characterized by the 2D Bloch vector ${\bf k}^{||}$. 
In the vacuum, far from the crystal surface, it is a superposition of 
plane waves: the plane wave propagating towards the crystal defines 
incident current, and the total current carried by the LEED state is 
the transmitted current. The ratio of the two currents is the 
transmission coefficient $T(E)$. In the bulk the LEED wave function
is a superposition of propagating (real $k^{\perp}$) and evanescent 
(complex $k^{\perp}$) Bloch waves.

This approach neglects inelastic scattering, so that in the case of
a non-zero transmission the LEED state necessarily includes a propagating 
Bloch wave. In what follows the propagating constituents are referred to 
as conducting states. Gross features of the $T(E)$ spectrum depend upon 
the band structure of the bulk crystal: the energies at which the band 
ceases transmitting the current (e.g. when the group velocity vanishes) 
and other critical points are reflected in experimental spectra. 
For example, $T(E)$ drops abruptly to zero when an energy gap in the 
${\bf k}^{||}$ projected band structure is encountered. However, 
owing to inelastic processes, one never observes zero transmission in 
the experiment, and instead of a sharp drop predicted by the simple 
theory one observes a rather smooth decrease of the transmission as a 
sign of the critical point.

In 1937 Slater \cite{sla} showed that the broadening of the spectral 
structures and the absence of energy gaps can be reproduced by adding an
imaginary term, the optical potential $-iV_{\rm i}$, to the potential
in the crystal half-space. Slater's idea is to associate the effect of
the optical potential with a spatial damping of the wave functions 
rather than with a decay in time. 
Then the Bloch vector acquires an imaginary part of the order of 
$V_{\rm i}/\hbar v^{\perp}$ ($v^{\perp}$ is the group velocity),
or, in terms of the mean free path $d=1/{\rm Im}k$, it is
$V_{\rm i} = \hbar v^{\perp}/d$. The optical potential is, 
thus, understood as the inverse lifetime of the quasi-particle. 
Alternatively, one can think of an electron with a lifetime 
$\tau=\hbar/V_{\rm i}$ and mean free path $1/{\rm Im}k$ that moves
in a real non-absorbing potential. \cite{StrocovExi}
In the present work we make an attempt to extract quantitative information 
on the energy dependence of  $V_{\rm i}$ from the measured $T(E)$ spectra by 
comparing the broadening of experimental and calculated spectral structures.

The unoccupied band structures of layered quasi-2D materials, such as 
graphite \cite{StrocovGraphi} or transition metal dichalcogenides,
\cite{StrocovTMDCs,Pehlke89,StrocovKluwer} are particularly interesting:
owing to the interlayer potential barrier, a nearly-free-electron model 
fails to provide even a qualitative picture of the unoccupied states.
The necessity to describe such states within an all-electron approach
for the crystal potential of general shape and to understand their relation 
to the diffraction process has been a driving force of the development of 
the Bloch-wave-based {\it ab initio} methodology. 
\cite{WACHUTKA86,HUMMEL_BROSS98,STILES_HAMANN88,WORTMANN_ISHIDA_BLUEGEL02} 
In this work, we use the extended linear augmented plane wave (ELAPW) 
$\mathbf{k}\cdot\mathbf{p}$ method, \cite{ikp} whose distinctive feature
is that it reduces the inverse band structure problem -- finding $k^{\perp}$
given ${\bf k}^{||}$ and $E$ -- to a matrix eigenvalue problem.

Of the layered materials, the band structure of 2H-NbSe$_{2}$ deserves
special attention. Its Fermi surface comprises two cylindrical sheets
and a tiny 3D electron pocket near the $\Gamma $ point,
\cite{corcoran,Straub99} all having different superconducting
properties. \cite{Yokoya01} Mapping of this pocket in the photoemission 
experiment is complicated by that its extension in $k^{\perp }$ is 
comparable with the final state $k^{\perp }$ broadening,
\cite{StrocovBessy,Rossnagel01} and it relies critically on the
knowledge of not only the final state dispersion but also lifetime. 

In this paper, we present experimental electron transmission spectra
of NbSe$_{2}$ and compare them to our {\it ab initio} calculations. 
The experimental technique is described in Sec.~\ref{experimental}. 
A brief account of our self-consistent band structure calculation is 
given in Sec.~\ref{bandstructure} and the computational method for LEED 
states is presented in Sec.~\ref{methodLEED}. The experimentally observed 
broadening of spectral structures is reproduced by including an imaginary 
part of the potential $V_{\rm i}$ into the Hamiltonian. 
In Sec.~\ref{broadening_analysis} we describe a procedure to obtain
quantitative information on the quasi-particle lifetimes from the
analysis of the shape of the experimental spectra. The observed energy
dependence of the optical potential is interpreted in terms of electron 
energy loss function. In Sec.~\ref{discussion} we present the unoccupied 
band structure of NbSe$_2$ in the $\Gamma$K direction and discuss possible 
limitations of the common description of the inelastic scattering by an 
optical potential. Optical properties are discussed in Sec.~\ref{optics}.

\section{EXPERIMENTAL} 
\label{experimental}
Essential details of our VLEED experimental technique are given elsewhere. 
\cite{StrocovCFS,StrocovRetField} Briefly, we used a standard 4-grid LEED 
optics operating in the retarding field mode. In this mode the
electrons are accelerated in the gun to the energies required to form
a well-focussed beam (normally from 100 to 300 eV) and then
decelerated in a retarding field between the gun and the sample.
This mode allows us to achieve the lowest primary energies without any 
significant degradation in focussing. However, angle dependent 
measurements are complicated by that the retarding field distorts the 
off-normal electron trajectories. To determine ${\bf K}^{\parallel}$, 
we parameterized its dependence on energy $E$ and the sample rotation 
angle $\alpha$ by a biquadratic function
$K^{\parallel}(E,\alpha )=\sum\limits_{l,m=0}^{2}A_{lm}E^{l}\alpha ^{m}$ 
with the coefficients $A_{lm}$ fixed by fitting to the experimental points 
with well-defined $K^{\parallel }$, namely to the angle dependent target 
current onsets, in which 
$K^{\parallel }$=$\sqrt{2m\left( E-e\phi \right) /\hbar }$, and to the
characteristic diffraction patterns when a diffracted beam hits the
electron gun exit, in which ${\bf K}^{\parallel }$ equals half the
surface reciprocal lattice vector ${\bf G}^{\parallel}/2$ 
(see Ref.~\onlinecite{StrocovRetField} for details).

The present experiment was carried out on an improved setup: Screening from 
the stray magnetic fields using $\mu $-metal shielding reduced the
displacement of the electron beam on the sample to less than 0.5 mm 
over the whole energy range. The position of the diffracted beams in the 
area obscured by the electron gun was controlled using a miniature 
fluorescent screen covered by ZnO:Zn low-energy phosphor, \cite{ThanksLevy} 
which was mounted at the gun exit. Careful adjustment of the
incidence angles was achieved using a custom made sample holder with 3
angular degrees of freedom. This is particularly important for the flaky
crystals of layered materials, whose surface after gluing to the sample
holder can appear a few degrees off. The tilt angle adjustment was coupled
to the linear shaft motion of our standard manipulator, and the azimuthal
angle was adjusted using an additional fork mounted on a rotary wobble
stick. This design guaranteed the incidence angle adjustment accuracy 
better than 0.25$^\circ$ in all angular degrees of freedom. 
A symmetric design of the sample holder prevented distortion of the
electron trajectories in the retarding field. The energy spread of the
primary beam was $\sim $0.25~eV HWHM. The size of the beam spot on the sample 
can be estimated as less than 0.5 mm through the whole experimental energy 
range.

As a practical matter, it should be noted that during the angle-dependent
measurements the electron spot can slightly displace along the surface due
to the retarding field and mechanical drifts in the manipulator. This
displacement can be a problem for the layered materials, whose crystals
closer to the periphery typically incorporate minor misoriented
crystallites. The sample position should therefore be optimized to avoid the
misoriented crystallites in the whole experimental energy and angle range.
This can be controlled most easily by inspecting the diffraction patterns on
the screen. In this respect the VLEED experimental setups based on the LEED 
electron optics have an advantage over the inverse photoemission based ones.

The experimental system provided two measurement modes: 1)
measurements of the elastic reflectivity into individual diffracted
beams, often referred to as the VLEED technique, using the LEED screen
and a CCD camera, and 2) measurements in the target current, commonly
referred to as the Target (or total, absorbed) Current Spectroscopy
(TCS) technique -- see, e.g., Ref.~\onlinecite{Komolov92}. The VLEED
mode gives most detailed information about the diffraction process,
but its principal limitation is that the reflected intensities in the
area obscured by the electron gun can not be measured (e.g. the
specular beam near the normal incidence). The TCS mode is free of this
problem and also benefits from greater experimental simplicity, but
its limitation is that the target current $I(E)$ gives the {\em total}
reflectivity which includes the elastic reflectivity $R(E)$ integrated
over all diffracted beams and, in addition, the inelastic reflectivity
$R_{\rm inel}(E)$ corresponding to the secondary electrons that leave 
the crystal. The inelastic contribution gives, however, only a rather
featureless background, so the structures in the $I(E)$ curves reflect
essentially the elastic electron transmission $T(E)=1-R(E)$, 
which contain the band structure information (the derivatives $dI/dE$ 
and $dT/dE$ are practically equivalent).

The term VLEED will further be used in reference to the dominant
physical mechanism forming the spectral structures in the individual
beams as well as in the total reflectivity, rather than to the
experimental technique. In this context we will refer to the target
current spectra also as the VLEED spectra.

Atomically clean (0001) surfaces of NbSe$_{2}$ were obtained by
standard cleavage in the vacuum chamber at a base pressure of
$7\times 10^{-10}$~mbar. Compared to other layered materials, the
lifetime of the NbSe$_{2}$ surface is small: in a day after cleavage
the spectral contrast significantly decreased and the background in
the LEED patterns increased, indicating strong surface contamination.

\section{Band Structure} 
\label{bandstructure}
The calculations are performed with the ELAPW $\mathbf{k}\cdot\mathbf{p}$ 
method. \cite{ikp,afc,new} The self-consistent potential was constructed 
within the local density approximation (LDA) of the density functional 
theory with the full-potential augmented Fourier components technique 
described in Ref.~\onlinecite{afc}. The basis set included 487 energy 
independent APWs (energy cutoff 10.2~Ry), and the extension of the radial 
basis set contributed another 200 basis functions. The extension was 
introduced following the prescriptions of Ref.~\onlinecite{new}.
The Brillouin zone (BZ) integrations were performed by the tetrahedron 
method with a mesh of 550 {\bf k} points that divides the irreducible BZ 
into 2187 tetrahedra.

All the occupied states down to the Nb 4$p$ semi-core band (at -31~eV) are 
treated as valence states. The density-of-states (DOS) function is in good 
agreement with the FLAPW calculation of Corcoran {\it et al}.~\cite{corcoran} 
and with X-ray photoemission measurements of Wertheim {\it et al}.~\cite{X2}
The five peaks due to the strong Se 4$p$ -- Nb 4$d$ hybridization have 
distinct counterparts in the experimental spectrum, see Fig.~\ref{xps}. 
The peak in the unoccupied DOS at 2.7~eV arises from localized states
of almost pure Nb~$d$ character. Thus, one can expect strong dipole-allowed
transitions from the Nb 4$p$ band to produce an absorption peak at 
$\hbar\omega\sim 34$~eV (see Sec.~\ref{optics}). We do not observe any 
localized states above $E-E_F=5$~eV; the complicated fine structure of
the DOS curve for higher energies reflects a non-free-electron-like 
behavior of the unoccupied states.

\section{Calculated VLEED spectra} 
\label{methodLEED}
Our calculations are based on the Bloch waves approach to the 
LEED problem, \cite{blochleed} in which the crystal is treated
as a semi-infinite system and the LEED wave function in the
crystal half-space is sought as a linear combination of propagating and 
evanescent solutions of the Schr\"odinger equation for a bulk crystal 
potential. Our implementation of the theory in the case of a singular 
all-electron crystal potential has been described elsewhere. \cite{ikp} 
Now we briefly sketch the computational procedure.
\subsection{Inverse Band Structure Problem} 
\label{methodIBS}
In the ELAPW-$\mathbf{k}\cdot\mathbf{p}$ method the complex band
structure can be obtained by an analytical continuation of the
Schr\"odinger equation to the complex ${\bf k}$ space. \cite{ikp} 
In application to semi-infinite systems, we use the 
ELAPW-$\mathbf{k}\cdot\mathbf{p}$ method to solve the 
{\em inverse band-structure problem}: given two real Cartesian 
components of the Bloch vector ${\bf k}^\parallel =(k_x,k_y)$ 
and the energy $E$, we find the values of $k^{\perp}$ that satisfy 
the Schr\"odinger equation
\begin{equation}
\label{sh}
\hat{H}\Psi (E,{\bf k}^\parallel + {\bf z}k^{\perp}_n;{\bf r})=
E\Psi(E,{\bf k}^\parallel + {\bf z}k^{\perp}_n;{\bf r})
\end{equation}
for the Bloch vector with a complex $z$-component $k^{\perp}_n$
(${\bf z}$ is a unity vector in the $z$-direction).

In the ${\bf k}\cdot{\bf p}$ method the wave function of a Bloch 
vector $({\bf k}^{\parallel}+ {\bf z}k^{\perp})$ is a product of
a trial function with the Bloch vector 
$({\bf k}^{\parallel}+ {\bf z}k^{\perp}_0)$ and the phase factor
$\exp[i(k^{\perp}-k^{\perp}_0) z]$. The trial function is a linear 
combination of the basis functions 
$\xi_j ({\bf k}_0;{\bf r})$ constructed at the reference point 
${\bf k}_0 = {\bf k}^{\parallel} + {\bf z}k^{\perp}_0$
with a real $k^{\perp}_0$:
\[
\Psi (E,{\bf k}^\parallel + {\bf z}k^{\perp};{\bf r}) = 
e^{\,i (k^{\perp}-k^{\perp}_0)  z}\,
\sum \limits_j
C_{j}^{\,k^{\perp}} (E,{\bf k}_0) \,\,\xi_j ({\bf k}_0;{\bf r}).
\]

In terms of the coefficient vectors 
$\vec{C}_n \equiv \{C_{j}^{\,k^{\perp}_n}\}$
the inverse problem can be written as a matrix equation
\begin{eqnarray}
\label{meq}
\left[\hat{H} + 2 \delta^{\perp}_n  \hat{P}^\perp +
({\delta^{\perp}_{n}}^{\,2} -E) \,\hat{O}\right]\vec{C}_n = 0 &,& \\
\nonumber
\delta^{\perp}_n  = k^{\perp}_n - k^{\perp}_0 &,&
\end{eqnarray}
with $\hat{H}$ being the Hamiltonian, $\hat{O}$ the overlap matrix, 
and $\hat{P}^\perp$ the $z$-projection of the momentum matrix. 
By introducing the vector 
${\vec{D}_n=-(2\hat{P}^\perp +\delta^{\perp}_n \hat{O})\,\vec{C}_n}$ 
we reduce Eq.~(\ref{meq}) to a linear eigenvalue problem of twice 
the dimension:
\begin{eqnarray}
\left( \begin{array}{cc}
0 & \hat{H}-E\hat{O} \\
\hat{I} & 2\hat{P}^{\perp}
\end{array} \right)
\left( \begin{array}{c}
\vec{D}_n \\
\vec{C}_n
\end{array} \right) =
\delta^{\perp}_n \left( \begin{array}{cc}
\hat{I} & 0 \\
0 & -\hat{O}
\end{array} \right)
\left( \begin{array}{c}
\vec{D}_n \\
\vec{C}_n
\end{array} \right).
\label{ikp}
\end{eqnarray}
Thus, the solutions of the inverse band-structure problem 
$k^{\perp}_n$ are obtained as the eigenvalues $\delta^{\perp}_n$ 
of this generalized non-Hermitian problem.

Because of the rather large unit cell of NbSe$_2$ (six atoms per unit 
cell) the solution of Eq.~(\ref{ikp}), especially for complex matrices, 
results in prohibitively time-consuming calculations. \cite{cmm1} It has 
turned out that the CPU time can be greatly reduced by orthogonalizing the 
original basis set, i.e., by seeking the trial function in terms of
the bulk Hamiltonian eigenfunctions $\psi_j ({\bf k}_0;{\bf r})$
at ${\bf k}_0$ rather than in terms of the basis functions 
$\xi_j ({\bf k}_0;{\bf r})$. Eliminating the overlap matrix alone 
makes the computing of Eq.~(\ref{ikp}) three times faster. The price for 
that is the necessity of recalculating the momentum matrix ${\hat P}^\perp$,
which scales quadratically with the number of both basis functions and
eigenfunctions. To further accelerate the procedure we cut the eigenvalue
spectrum and retain only $\psi_j$ with energies lower than 5~Ryd above
the spectral interval we are interested in. A suitable cutoff energy is
difficult to determine {\it a priori}, however, the quality of the results
can be verified by comparing the resulting inverse band structure 
$k^{\perp}(E)$ with the usual band structure $E(k^{\perp})$.
\subsection{Constructing the LEED Function} 
In the bulk half-space the LEED function $\Phi$ is expanded in terms of the 
solutions 
$\Psi_n\equiv\Psi(E,{\bf k}^\parallel+{\bf z}k^{\perp}_n;{\bf r})$
of Eq.~(\ref{ikp}). The quality of the wave functions $\Psi_n$ is known to 
deteriorate with growing $|\delta^{\perp}_n|$. \cite{ikp} With the present 
computational setup reliable solutions were found to be restricted to the 
interval ${\rm Im}\, \delta^{\perp}_n < 3$~\AA$^{-1}$. To construct the LEED 
function, we take all the solutions in this interval, which for NbSe$_2$ is 
typically ten (at the vacuum level $E_{\rm vac}$) to twenty Bloch waves 
(at 40~eV above $E_{\rm vac}$). Then a Laue representation of the functions 
is constructed
\begin{equation}
\label{val}
\Psi_n ({\bf r^{||}},z) = \sum \limits_{s=0}^{N_F-1} f_{sn}(z)
\exp\,[\,i(\,{\bf k}^{||} + {\bf G}^{||}_s\,)\,{\bf r^{||}}\,].
\end{equation}
Here ${\bf r^{||}} = (r_x,r_y)$ and ${\bf G}^{||}_s$ are surface 
reciprocal vectors. To construct the Laue representation we use a rapidly 
converging 3D Fourier decomposition of the wave function, which differs from 
the true all-electron wave function only in a close vicinity of the nucleus 
(see the gouging technique described in Refs.~\onlinecite{afc} and 
\onlinecite{lfe}).
Each surface Fourier component $f_{sn}(z)$ is smoothly continued into the 
vacuum half-space by a linear combination of two plane waves, one of which 
is the outgoing vacuum solution $\exp\,[\,ik_sz\,]$ with 
$|\,{\bf k}^{||} + {\bf G}^{||}_s\,|^2 + k_s^2 = E$,
and the other one is a decaying (unless ${\bf G}^{||}_s=0$) 
plane wave $\exp\,[\,iq_sz\,]$ with a purely imaginary $z$-component 
$q_s$. In the present calculation it has been chosen $k_s^2-q_s^2 = \Delta E$ 
for all $s$ with $\Delta E=3$~Ryd. For ${\bf G}^{||}_0=0$
one of the waves represents the incident electron wave, and
the other one the reflected wave, i.e., $q_0=-k_0$.

Thereby the functions $\Psi_n$ are defined in the whole space and they 
are smoothly continuous everywhere. The LEED function $\Phi$ is sought 
as a linear combination of the functions that minimizes the value
\begin{equation}
\label{tomin}
\| (\hat H-E)\Phi \| = \int \limits_{\rm vacuum} 
|(\hat H-E)\Phi({\bf r})|^2 \, d{\bf r},
\end{equation}
under the constraint that the incident current is equal to unity. The
resulting function $\Phi({\bf r})$ is the closest to a solution of
the Schr\"odinger equation in the surface region. By construction the
trial function satisfies the Schr\"odinger equation both in the bulk
half-space and far from the surface in the vacuum, so that only a
finite slab in the vicinity of the surface contributes to the integral
(\ref{tomin}). This slab may include the part of the crystal where the
potential strongly deviates from the bulk potential, but in this
calculation we adopt a simplified step-like shape of the surface
barrier, so the potential changes abruptly in the middle of the
van-der-Waals gap from its bulk distribution to the constant vacuum
value. The generalization of the method to the case of an arbitrary
potential in the interface region has been described in
Ref.~\onlinecite{vam}.

The minimization procedure $\delta \| (\hat H-E)\Phi \| = 0$ leads
to a system of linear equations. If the auxiliary tails $\exp\,[\,iq_sz\,]$ 
accidentally cancel then the result is equivalent to the simple matching of 
wave functions. It is, however, difficult to predict whether the matching is 
possible and to separate the matching error from the errors introduced at the
stage of obtaining the LEED function constituents $\Psi_n$. \cite{ikp}
If the exact matching is impossible the variational method yields a solution 
that is smoothly continuous by construction -- this function is thought to be 
the best approximation for the LEED function we seek. The accuracy of the 
ultimate result can be estimated {\it a posteriori} by checking the current 
conservation in the LEED function, i.e., by comparing the current transmitted
into the crystal by the superposition of the propagating Bloch waves $\Psi_n$
to the current in the vacuum carried by the superposition of plane waves.
\subsection{Normal Incidence Spectrum}
The ${\bf k}^{||}=0$ spectra along with the underlying real band
structure are presented in Fig.~\ref{cb}. The $V_{\rm i}=0$ calculation 
yields partial currents carried by the Bloch states (shown by whiskers).
It  reveals the conducting states responsible for the transmission of the 
current, thereby offering an interpretation of the VLEED spectrum in terms 
of conducting fragments of the band structure: the abrupt changes in 
$T(E)$ all reflect critical points in the conducting bands.

The deviation of the $k^{\perp}(E)$ points by the inverse 
ELAPW-$\mathbf{k}\cdot\mathbf{p}$ from the $E(k^{\perp})$ lines by the
direct method with the full basis is seen to be negligible. 
Together with the reasonably small current non-conservation this justifies 
the basis set reduction we have undertaken to make the calculations feasible.

A special feature of the normal incidence spectrum is that the conducting 
bands do not overlap, i.e., there is only one dominant propagating 
constituent in the LEED function. In this respect the picture is similar 
to that of the Slater 1D model. \cite{sla} The inclusion of the imaginary 
term lifts the principal difference between the propagating and evanescent 
waves (whose properties to couple to the incoming wave may be very similar) 
and thereby levels out the intensity variations (see the dashed curve in 
Fig.~\ref{cb}). In particular, already at a moderate value of 
$V_{\rm i} = 0.5$~eV the narrow gaps in the $T(E)$ spectrum almost 
completely disappear, so one can hardly expect them to be observed 
in the experiment. Wider gaps remain well visible, although the reflected 
intensities are strongly reduced and the structures broadened with respect 
to those predicted by a Hermitian Hamiltonian. In the next section we will 
use these properties of the Slater theory to extract quantitative information 
about the damping of quasi-particles from the shape of experimental spectra.

\section{Determining the energy dependence of $V_{\rm \lowercase{i}}$} 
\label{broadening_analysis}
We shall now compare our measured target current spectra $I(E)$ with the
calculated transmission coefficient $T(E)$. First of all, we need to 
bring the $I(E)$ spectrum to the same  absolute units as $T(E)$. This 
is not trivial because of the presence of the unknown background 
caused by the secondary electrons contributing to the reflectivity. 
In view of the close similarity of the fine structure of the measured 
$I(E)$ and theoretical $T^{\,\rm th}(E)$ spectra over a wide energy region, 
see Fig.~\ref{ni}, we have chosen to determine the {\it experimental}
$T(E)$ curve by fitting $I(E)$ to $T^{\,\rm th}(E)$ with the linear
transformation $\,T^{\,\rm exp}(E) = aI(E) + b + cE$. Here the
function $b + cE$ represents a linearly varying background.

Strong intensity variations in the $T(E)$ spectrum are much more
important for the band structure information than the shape of the
maxima,\cite{StrocovCFS} which are very broad even if the damping is
neglected (see Fig.~\ref{cb}).  Thus, for a conclusive comparison, the
extrema of the $dT(E)/dE$ function must be analyzed.

Owing to computational instabilities, theoretical $T(E)$ curves
contain numerical noise, which should be excluded before the
derivative is calculated.  To remove the jaggedness we average the
$T(E)$ function within an interval of width $\Delta E$ by fitting it
to a parabola: for a given $E$ a parabola is constructed that
approximates the raw $T(E)$ spectrum over the interval 
$[E-\Delta E/2;E+\Delta E/2]$ and its value at $E$ is taken to be the 
intensity of the smoothed $T(E)$ spectrum. The width $\Delta E$ of the 
interval is 0.6~eV at the vacuum level and it grows linearly up to 2.4~eV 
at 50~eV above $E_{\rm vac}$. Thereby slow changes of intensity are not
affected and only the noise is removed.

We extract the energy dependence of the optical potential $V_{\rm i}(E)$ from
the experimental spectra by comparing the sharpness of the $dT/dE$ extrema 
(maxima and minima) in the experimental and theoretical spectra.
To quantitatively characterize the sharpness, we associate the function 
$dT/dE$ in the interval between its two zeros $E_1$ and $E_2$ with a parabola 
that passes through zero at $E_1$ and $E_2$ and embraces the same area $S$ as 
the $dT/dE$ function, $S=T(E_2)-T(E_1)$. Then the curvature of the parabola 
$6S/(E_2-E_1)^3$ is taken to express the sharpness of the structure. This 
parameter changes very strongly over the spectrum, so in Fig.~\ref{si} we 
show the cube root of the curvature $\sqrt[3]{S}/(E_2-E_1)$.
We have performed a series of calculations for the normal incidence with an 
energy independent $V_{\rm i}$ ranging from 0.5 to 3~eV and considered the 
sharpness of each structure as a function of optical potential. Three examples
are shown in the right panel of Fig.~\ref{si}. 
The value of sharpness from the left panel is taken to yield the value of
the optical potential in the right panel. One immediately notices a sharp
increase of the optical potential between 16 and 21~eV: the two minima change 
their shape very similarly with increasing the $V_{\rm i}$ (circles and 
diamonds in right panel of Fig.~\ref{si}) and the minimum at 21~eV in the 
experiment is significantly broader than the one at 16~eV (see left panel of 
Fig.~\ref{si}).

The values of $V_{\rm i}$ determined in this way are necessarily
overestimated. First, in addition to the inelastic processes in the
electronic system, there exist broadening mechanisms dependent upon
experimental conditions, such as surface roughness and finite energy
resolution. Also the angular spread of the incident beam contributes
to the broadening owing to the ${\bf k}^{\parallel }$ dispersion of
spectral structures. Secondly, our theoretical spectra do not take
into account inelastic scattering in the surface barrier region, e.g.,
scattering on the surface defects. This effect results in an additional 
reduction of the spectral structures but, unlike the inelastic scattering 
in the bulk, it hardly affects their energy broadening.  These mechanisms 
should, however, be less significant for the layered materials because of 
the rather small electron density at the surface and a small concentration 
of defects on the cleaved surface. 
On the other hand, our absolute values of $V_{\rm i}$ are in accord 
with the recent {\it ab initio} results on the quasi-particle lifetimes
for noble and transition metals: \cite{lifetimes} For energies around
$E-E_F=5$~eV the calculated lifetimes do not exceed a few fs 
($V_{\rm i}\sim 0.2--0.7$~eV).

The energy dependence $V_{\rm i}(E)$ derived from the experiment is
presented in Fig.~\ref{vo}. To get an idea of how reliable the obtained 
$V_{\rm i}(E)$ dependence is we compare the normal incidence data (circles) 
with the data for ${\bf k}^{||}=\frac{1}{4}\Gamma{\rm K}$ 
(${\Large\times}$). (The off-normal incidence spectra are shown in 
Fig.~\ref{gkt}; $\frac{1}{4}|\Gamma{\rm K}| = 0.3$~\AA). 
The absolute values of $V_{\rm i}$ 
agree very well and both spectra suggest a sharp increase at around 20~eV. 
The non-steady behavior of the ${\bf k}^{||}=\frac{1}{4}\Gamma{\rm K}$ 
derived values around 30~eV seems to have no physical meaning; it reflects 
the disagreement between theory and experiment in shape of the $T(E)$
maximum at 25~eV, which in section \ref{discussion} we explain by the
deviation of the calculated band structure from the experimentally
observed one.

An optical potential is associated with the imaginary part of the
electron self-energy $\Sigma$, whose energy dependence is expected to
reflect singularities of the energy loss function 
$-{\rm Im}[1/\varepsilon({\bf q},\omega)]$. In particular, in the 
$GW$ approximation, \cite{hedin_lundqvist69} Im$\,\Sigma$ is given by 
an integral, whose integrand contains the inverse dielectric function (DF). 
We are not in a position to calculate the lifetime of the LEED constituents 
{\it ab initio}. A rough idea of the average effect of plasmon excitations 
on the energy dependence of the electron lifetime can be obtained from
the DF for ${\bf q}=0$ by plotting the integral
$\int\limits_0^E\,-{\rm Im}[1/\varepsilon({\bf q}=0,\omega)]\,\,d\omega$.
We have calculated the DF in the random phase approximation, see
Sec.~\ref{optics}. In Fig.~\ref{vo} we compare the integral function
to the experimental dependence $V_{\rm i}(E)$. The sharp increase of
the optical potential agrees well with the plasmon location at 21~eV
(see also Fig.~\ref{op} in Sec.~\ref{optics}), which is confirmed by
the transmission energy loss measurements by Bell and Liang. \cite{O2}
Our theoretical DF shows also a second plasmon peak at 35~eV, which
may be responsible for the step-like growth of the optical potential
at higher energies.

\section{$\Gamma$K direction} 
\label{discussion}
Experimental and theoretical $T(E)$ curves for ${\bf k}^{||}$ ranging
from zero to $\Gamma$K are presented in Fig.~\ref{gkt}. The experimental and 
theoretical spectral structures show a good agreement in both energy 
dispersion and relative amplitudes. Owing to the increase of the incident 
beam spread in ${\bf k}^{\parallel }$, the low-energy part of the experimental
spectra at large ${\bf k}^{\parallel }$ looks smeared compared to the theory. 
Note that because of the vector $\mathbf k^{||}$ changing with the incident 
electron energy the spectra cannot be compared in a one-to-one manner. To 
facilitate the comparison of the dispersion of the spectral structures we 
present a gray-scale $T(\mathbf k^{||},E)$ plot, Fig.~\ref{gs_gki}. A very 
good agreement between the measured and calculated $\mathbf k^{||}$ dispersion
of the $T(E)$ intensity proves that the theoretical approach we have adopted 
is adequate. The results have turned out rather sensitive to details of the 
potential distribution in the bulk half-space; we stress that the advantage 
of the ELAPW-$\mathbf{k}\cdot\mathbf{p}$ method of accurately taking into 
account the strong non-muffin-tin effects makes the method an indispensable 
tool for studying the electron diffraction on layered materials.

We now analyze the information on the bulk band structure contained in our 
experimental data (and deliberately blurred by the optical potential in 
the theory). The critical points in the $k^{\perp }$ dispersion of the 
conducting bands (e.g. the band edges) manifest themselves as the extrema 
in the $dT/dE$ curve. Of course, reflected is the optical-potential-damped 
(completely complex) band structure, in which the band dispersions are 
smoothed, and the critical points are shifted by some tenth of eV from their 
$V_{\rm i}=0$ positions. \cite{StrocovExi} The deviation of the surface 
barrier from the step-like form may also cause shifts of $dT/dE$ extrema, 
but for the layered material this effect seems to be insignificant.

To stress the ${\bf k}^{\parallel }$ dispersion of the spectral structures
we construct a gray-scale plot, Fig.~\ref{gs_gkd}, in which the shading shows 
the energy area covered by the conducting bands (the regions between $dT/dE$ 
maxima and minima); ${\bf k}^{||}$ projected band gaps then appear as 
white areas. Again one observes excellent overall agreement between the 
experiment and theory. The plot, however, reveals also differences in finer 
details, which reflects fundamental limitations of the theory employing a 
one-electron band structure and optical potential. The deviations are 
necessarily present due to our neglect of the real part of the self-energy
$\Sigma(E,{\bf k})$ -- it is replaced by an exchange-correlation potential in 
the local density approximation. For example, the white area ranging from 
${\bf k}^{||}=0$, $E-E_F=27$~eV to ${\bf k}^{||}=0.7$~\AA$^{-1}$, 
31~eV in the experiment is shifted towards higher energies by about 1~eV 
with respect to its calculated counterpart. In particular, this causes the 
higher width of the measured $T(E)$ maxima between 25 and 29~eV than in the 
theory (see the discussion in Sec.~\ref{broadening_analysis}). The difference is
especially well visible between 0.2 and 0.4~\AA$^{-1}$. The discrepancies 
are seen to significantly increase above $\sim $40 eV: an overall shift of 
the experimental bands to higher energies is observed.

The deviations are seen to increase at larger ${\bf k}^{\parallel}$.
From the band-structure point of view the main difference between the normal 
and off-normal incidence spectra is that in the latter case the conducting 
bands intersect; in other words the LEED state contains more than one 
propagating Bloch constituent.
An example for ${\bf k}^{\parallel}=\Gamma$K is shown in Fig.~\ref{kh}, 
lower panel. The theoretical $T(E)$ curve for negligible absorption (solid 
line) has a different character from the normal incidence spectrum 
(Fig.~\ref{cb}): there are no energy gaps in the band structure along 
the KH line, and the transmission never drops to zero and never reaches 
unity -- it almost never exceeds 0.8. Thus, not only minima but also maxima 
are affected by the optical potential. Also the positions of spectral minima 
strongly depend upon the function $V_{\rm i}(E)$. For example, with increasing
$V_{\rm i}$, the minimum at 19.6~eV moves away from its measured location 
towards lower energies and in the spectrum calculated with the energy 
dependent $V_{\rm i}$ it appears at 17.9~eV. Such a high sensitivity 
of results to a phenomenological parameter may be an important limitation 
of the theory. Although the overall shape of the 
${\bf k}^{\parallel}=\Gamma$K spectrum is well 
reproduced by the theory, the band structure information may be 
incorrectly transferred to the $T(E)$ curve. 

Presumably, the weak point is the presence of several propagating Bloch waves 
in the band-structure decomposition of the LEED function. 
In connecting the optical potential to the electron attenuation we basically 
rely on the relation Im$\,k^{\perp} \sim V_{\rm i}/v^{\perp}$. The parameter 
$V_{\rm i}$ is, thus, easy to operate when there is only one wave, but its 
meaning is not that transparent when there are several waves with different 
velocities (which, in addition, are connected by the matching conditions). 
This point of view is supported by Fig.~\ref{ds} in which we compare the 
measured ${\bf k}^{\parallel}$ dispersion of the $T(E)$ minima with 
theoretical results for the energy dependent $V_{\rm i}$ and for a constant 
moderately small $V_{\rm i}$ of 0.5~eV. The latter results are expected to 
provide undistorted information on the band structure. One can see that 
whenever there is a disagreement between small $V_{\rm i}$ and large 
$V_{\rm i}$ results, for small $k^{\parallel}$ the experimentally determined 
function $V_{\rm i}(E)$ brings the positions of the minima closer to the 
experiment (areas A and B). On the contrary, for large $k^{\parallel}$ the 
$V_{\rm i}=0.5$~eV results are often closer to the experiment (areas C and D).

Such deviations can be corrected by introducing a position dependent (in
general non-local) $V_{\rm i}$, which would allow to control the behavior 
of the complicated LEED state. In this case different Bloch constituents 
would be damped differently depending on their spatial distributions.
Such an approach within the multiple-scattering formalism of VLEED has 
been developed by Barto{\v s} {\it et al.} in Ref.~\onlinecite{viofr}.

\section{Optical Properties} 
\label{optics}
Optical properties of the crystal provide independent information on the
unoccupied band structure. Within the one-particle theory -- random phase 
approximation \cite{rpa} without local field effects -- the calculations
employ the same unoccupied bulk states as the LEED calculations of 
Sec.~\ref{methodIBS}. 

The imaginary part of the dielectric function $\varepsilon_2(\omega)$ 
was calculated as an integral over the BZ with the tetrahedron method
(for computational parameters see Sec.~\ref{bandstructure}). The real 
part $\varepsilon_1(\omega)$ was determined out of $\varepsilon_2(\omega)$ 
by the Kramers-Kronig integration with the energy cutoff 
$\hbar\omega_{\,\rm max}=70$~eV [The values of $\varepsilon_2(\omega)$ 
above 38~eV are underestimated because of the finite number of bands taken 
into account -- from Nb 4$p$ (-31~eV) up to 38~eV above the Fermi level.]

In Fig.~\ref{op} we compare our {\it ab initio} results with the 
reflectivity measurements by Liang \cite{O1} and transmission energy 
loss spectrum by Bell and Liang. \cite{O2} The experimental data are 
in excellent agreement with the calculations. The main plasmon peak 
at 21~eV corresponds to the termination of transitions from the 4$d$ 
valence band states. Our calculations predict also a rather strong peak
at 36~eV due to the transitions from the semi-core Nb 4$p$ states to 
the unoccupied Nb 4$d$ peak at $E-E_F = 3$~eV (see Fig.~\ref{xps}).
The high-energy plasmon is important to understand the energy dependence
of the electron lifetime.

\section{ conclusions }
Angle-dependent VLEED measurements in the target current mode have been 
performed on the layered 2H-NbSe$_{2}$. Tight control over the electron 
beam and adjustment of the incidence angle in three angular degrees of 
freedom enabled highly accurate measurements for the incidence wave
vector ${\bf k}^{\parallel }$ scanning the $\Gamma$K line of the
Brillouin zone.
Based on the self-consistent band structure, {\it ab initio} VLEED spectra 
have been calculated with the ELAPW-$\mathbf{k}\cdot\mathbf{p}$ method, the 
inelastic scattering being modelled via the optical potential.

A novel methodological aspect is the algorithm to extract the information on 
the electron lifetime from the VLEED experiment. The obtained results are 
consistent with available experimental and theoretical information on the 
dielectric function of NbSe$_2$.

Owing to the complicated non-free-electron-like unoccupied band structure
of NbSe$_2$, the VLEED spectra show rich structure over the energy region 
up to 50~eV above the Fermi level. A good agreement between experiment and 
theory in the energy location of the TCS structures over the whole interval
makes it possible to interpret the observed shape of the spectra in terms
of the energy dependent optical potential.

The comparison of the experimental and theoretical results focusses on the
surface-projected band structure represented by the ${\bf k}^{\parallel }$ 
dispersion of the extrema in the energy derivative of the transmission 
coefficient $dT/dE$. A good agreement between the experiment and theory 
proves that our theoretical approach is capable of accurate description 
of the unoccupied bands of NbSe$_{2}$. Minor discrepancies have been observed,
which are traced back to the imperfectness of our one-electron approach
as well as to intrinsic shortcomings of the phenomenological treatment of 
inelastic scattering.

\section*{Acknowledgments} 
The authors benefited from discussions with I.~Barto{\v s}. We acknowledge 
the support of Deutsche Forschungsgemeinschaft to E.E.K. (Forschergruppe DE 
412/21), and to V.N.S. and R.C. (CL124/5-1).


\begin{figure} 
\caption{\label{xps} Density of states for NbSe$_2$: total DOS
(solid line) and $l$-projected Nb $d$ (dot-dashed line) and Se $p$ 
(dotted line) partial DOS per atom. (Note the energy scale break at 
7~eV.) Dashed curve shows the X-ray photoemission 
spectrum by Wertheim {\it et al}.~\protect\cite{X2} 
}  
\end{figure}             

\begin{figure} 
\caption{\label{cb} 
Calculated unoccupied band structure of NbSe$_2$ in the $\Gamma$A direction.  
The lines in the left panel show the band structure $E(k^{\perp})$
for $V_{\rm i}=0$ obtained with the direct ELAPW-$\mathbf{k}\cdot\mathbf{p}$ 
method (without the basis reduction). The conducting bands are marked by
whiskers, whose upper ends show the band structure $k^{\perp}(E)$ obtained 
with the inverse method. The length of the whisker is proportional to the 
contribution of the Bloch wave to the target current. 
Transmission coefficient $T(E)$ (calculated in the vacuum half-space) is 
shown in the right panel by the solid line. The shaded area shows its
deviation from the current in the bulk (the sum of the group velocities
times the weights with which the constituents enter the LEED function). 
The dashed line is the $T(E)$ spectrum with the optical potential 
$V_{\rm i} = 0.5$~eV added to the Hamiltonian $\hat H$ in 
Eq.~({\protect\ref{ikp}}).
}
\end{figure}                      

\begin{figure} 
\caption{\label{ni}
Comparison of the normal incidence target current $I(E)$ and 
transmission coefficient $T(E)$ spectra of NbSe$_2$. The measured 
$I(E)$ curve is shown by the dot-dashed line. Theoretical $T(E)$ and
$dT/dE$ spectra are shown by dashed lines in the lower and 
upper panels respectively; they are calculated with an energy
dependent optical potential [the $V_{\rm i}(E)$ function is shown 
in Fig.~\ref{vo}]. To arrive at the experimental $T(E)$, 
the original $I(E)$ function is fitted to the theoretical $T(E)$ 
by a linear transformation that involves scaling and subtracting 
the background. The resulting experimental $T(E)$ and $dT/dE$ 
curves are shown by solid lines. 
}   
\end{figure}                       

\begin{figure} 
\caption{\label{si}
Left panel: sharpness of the extrema in the experimental ($\times$) and 
theoretical ($\circ$) $dT/dE$ spectra (see upper panel of Fig.~\ref{ni}).
Negative values are ascribed to minima and positive to maxima.
Right panel: dependence of the sharpness upon the optical potential for two 
minima (at 16 and 21~eV) and a maximum (at 30~eV) in the $dT/dE$ spectrum.
}   
\end{figure}                       

\begin{figure} 
\caption{\label{vo}
Dependence of the optical potential $V_{\rm i}$ on the incident electron 
energy extracted from the normal ($\Huge\circ$) and an off-normal 
(${\Large\times}$) incidence spectrum. The function $V_{\rm i}(E)$ used in 
the calculations is shown by the dashed line. The solid line is the function 
$\int\limits_0^E\,-{\rm Im}[1/\varepsilon({\bf q}=0,\omega)]\,\,d\omega$.
}   
\end{figure}             

\begin{figure} 
\caption{\label{gkt}
Experimental (left) and theoretical (right) angular dependence of the
$T(E)$ spectra with the vector ${\bf k}^{||}$ scanning the $\Gamma$K
line of the BZ. In the experiment the sample rotation angle $\alpha$
increases in steps of 0.5$^\circ$. The onsets of the spectra are at the 
energy $e\varphi + (\hbar ^{2}/2m)K_{\parallel }^{2}$ at which the electrons 
start penetrating into the solid. Owing to the retarding field, the
incidence angle differs from $\alpha$ and depends on energy in such a way 
that ${\bf k}^{\parallel }$ varies along the spectrum almost linearly. 
In the left panel the ${\bf k}^{\parallel }$ values are indicated that
correspond to the onset of the spectrum and to $E-E_F=50$~eV. 
Note the scale change at 18.5~eV (marked by a dashed line).
}   
\end{figure}                      

\begin{figure} 
\caption{\label{gs_gki}
Experimental (left) and theoretical (right) ${\bf k}^{\parallel }$ 
distribution of the $T(E)$ spectral intensity shown in a linear gray scale.
White areas correspond to maximal $T(E)$. [Original $T(E)$ curves are shown 
in Fig.~\ref{gkt}]. The energy region shown begins 0.5~eV above the 
transmission onset. Note the scale change at 18.5~eV (marked by a white line).
}
\end{figure}                        

\begin{figure} 
\caption{\label{gs_gkd}
Experimental (left) and theoretical (right) ${\bf k}^{\parallel }$ 
dispersion of the $dT/dE$ spectra. [Original $T(E)$ curves are shown 
in Fig.~\ref{gkt}]. The shading fills the regions between the maxima and 
minima in the $dT(E)/dE$ spectra, i.e., the area where the second 
derivative $d^{2}T/dE^{2}$ is negative. Physically, the shaded area 
shows the surface-projected dispersion $E({\bf k}^{\parallel})$ of 
the conducting bands. The gray scale shows the amplitude of $d^{2}T/dE^{2}$ 
(in a logarithmic scale), which characterizes the amplitude and sharpness of 
the extrema. The shown energy region begins 0.5~eV above the transmission 
onset.
}
\end{figure}                      

\begin{figure} 
\caption{\label{kh}
Dependence of the electron transmission spectrum for 
${\bf k}^{\parallel}=\Gamma$K on the optical potential. Experimental 
spectrum is shown by the dashed line. In the lower panel conducting fragments
of the band structure in the H-K-H interval are shown (see also the 
caption of Fig.~\ref{cb})  
}   
\end{figure}                       

\begin{figure} 
\caption{\label{ds}
${\bf k}^{\parallel}$ dispersion of $T(E)$ minima: experiment
(dots); theory with the energy dependent $V_{\rm i}$ (circles)
and with $V_{\rm i}=0.5$~eV (lines).
}   
\end{figure}                       

\begin{figure} 
\caption{\label{op}
Optical properties of NbSe$_2$ for two light polarizations: 
$\mathbf E\perp c$ (solid lines) and $\mathbf E\parallel c$ 
(dashed lines). The experimental reflectivity spectrum \protect\cite{O1} 
for $\mathbf E\perp c$ is shown in the upper panel by the dot-dashed line. 
The triangles in the lower panel show the energy location and intensity of 
the low-energy structures of the measured \protect\cite{O2} EELS spectrum. 
The fragment of the Lorentzian function (dotted line) reproduces the 
experimentally observed \protect\cite{O2} energy location, intensity, 
and width of the main plasmon.
}
\end{figure}                      

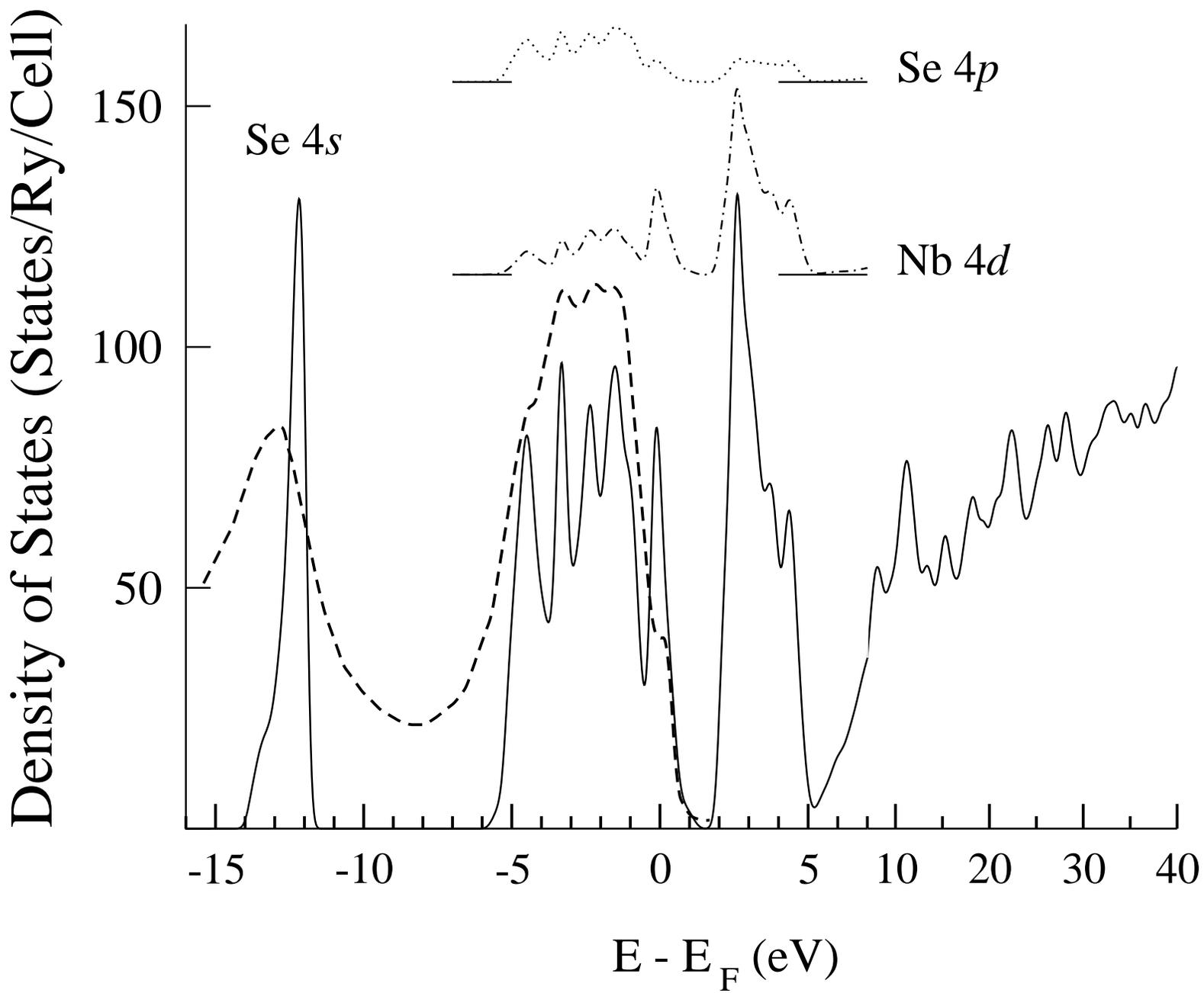
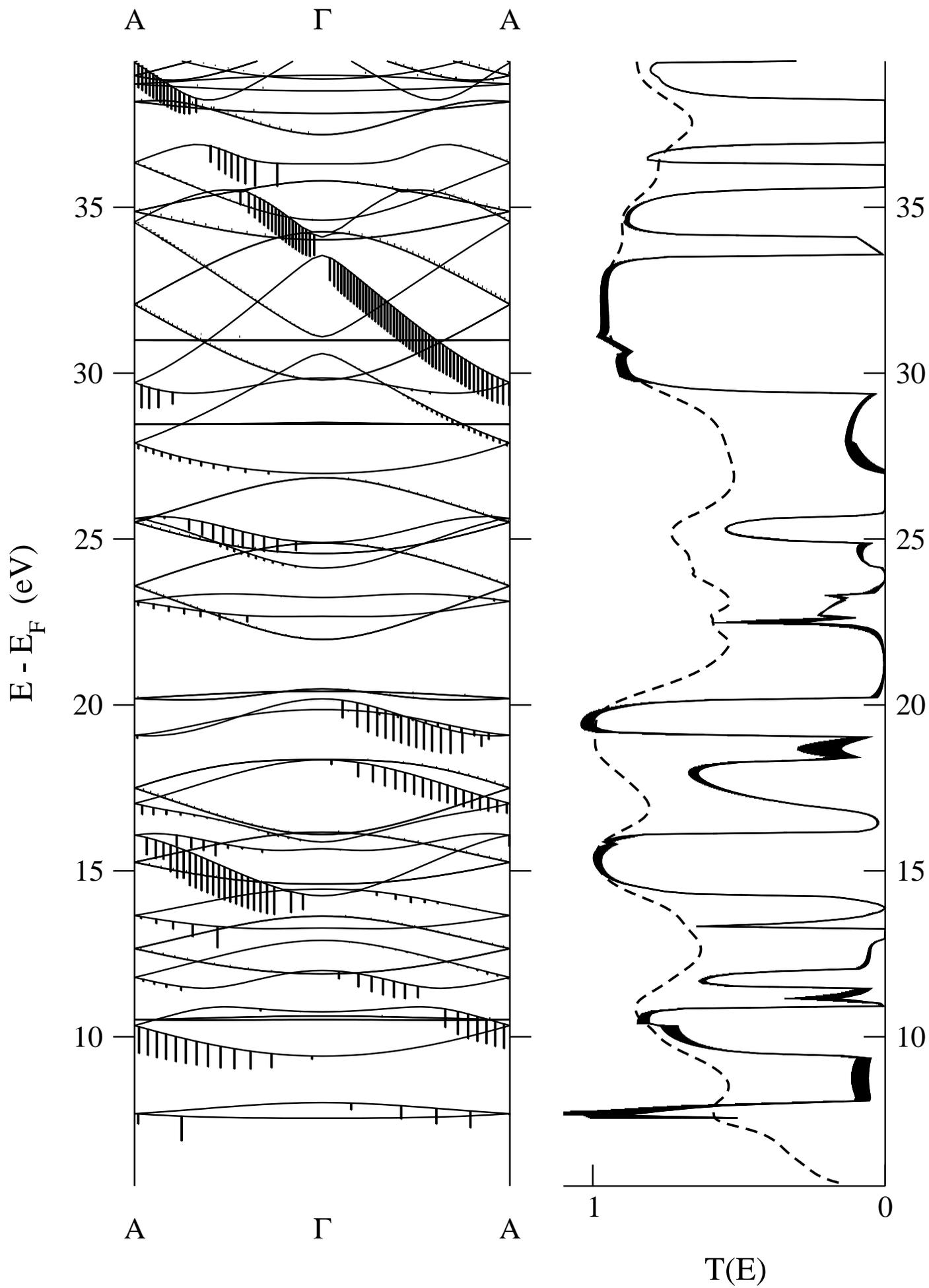
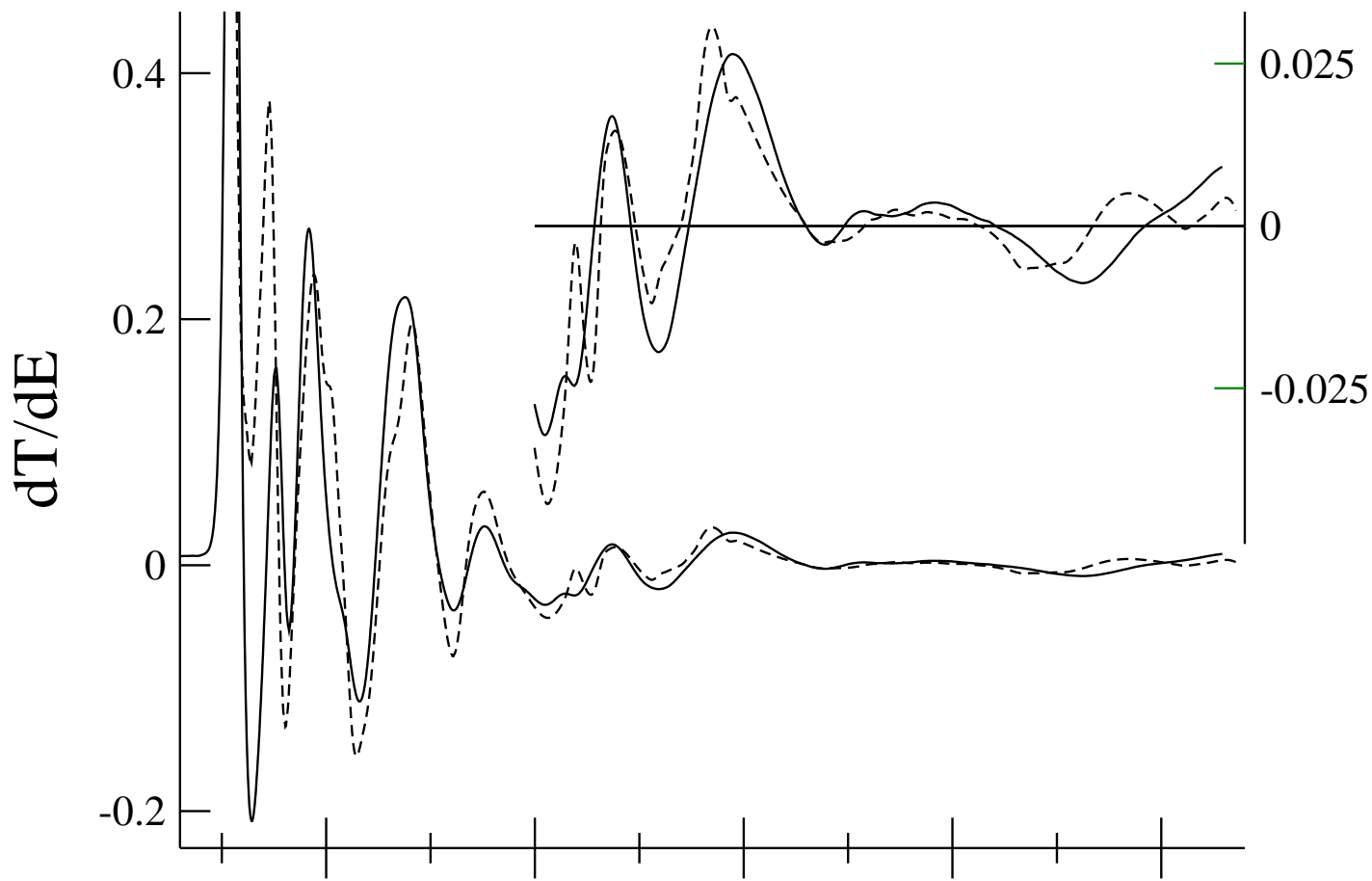
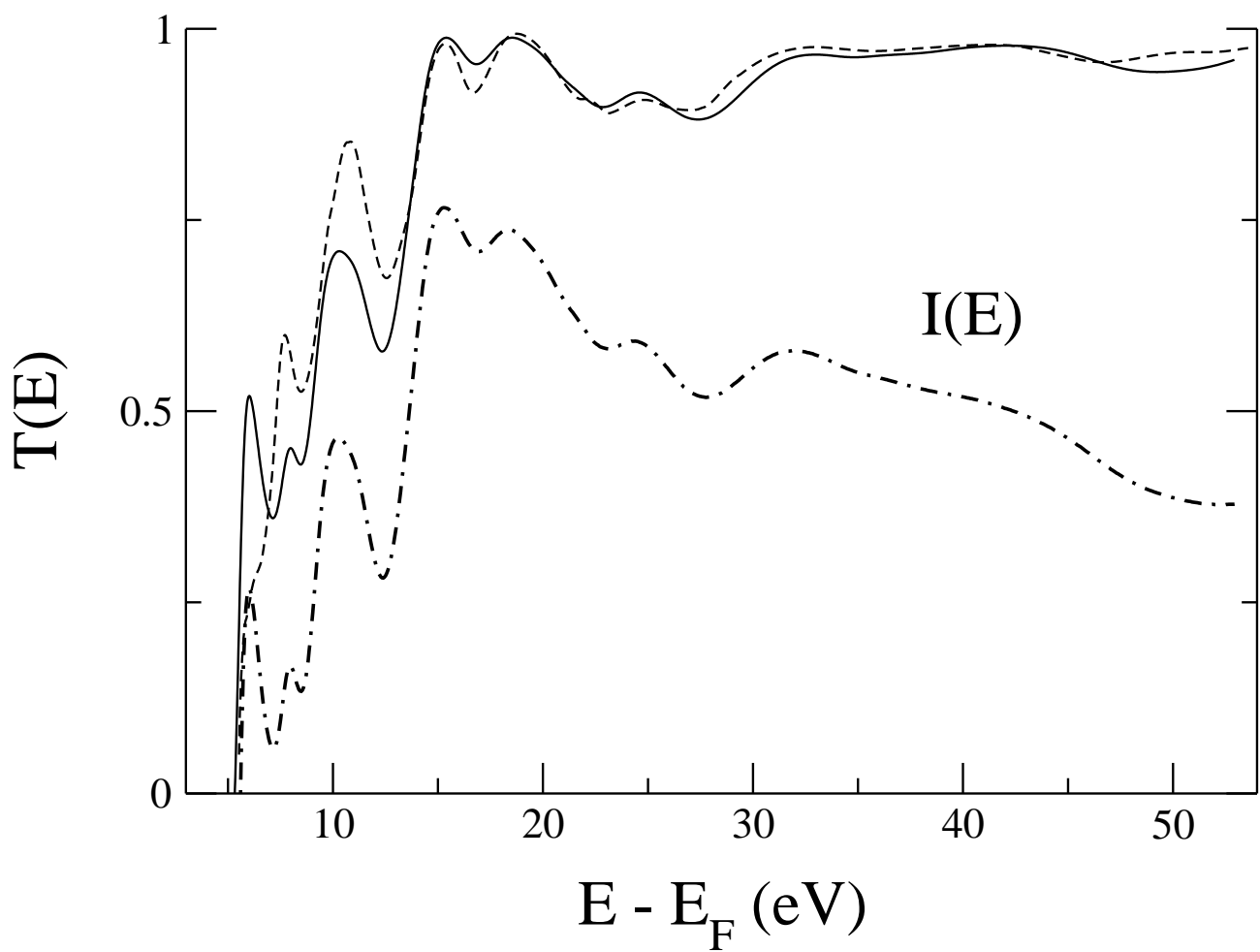
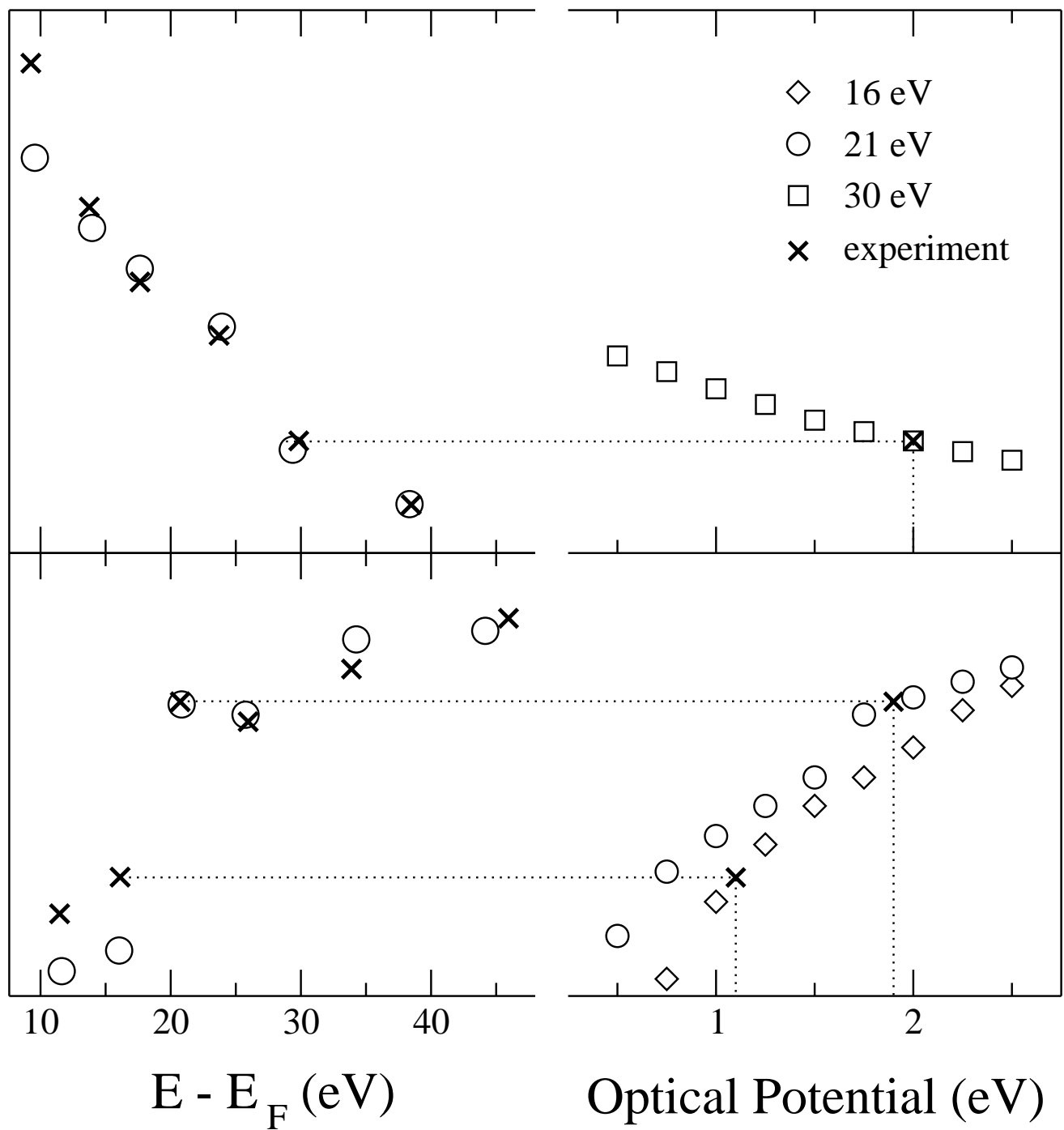
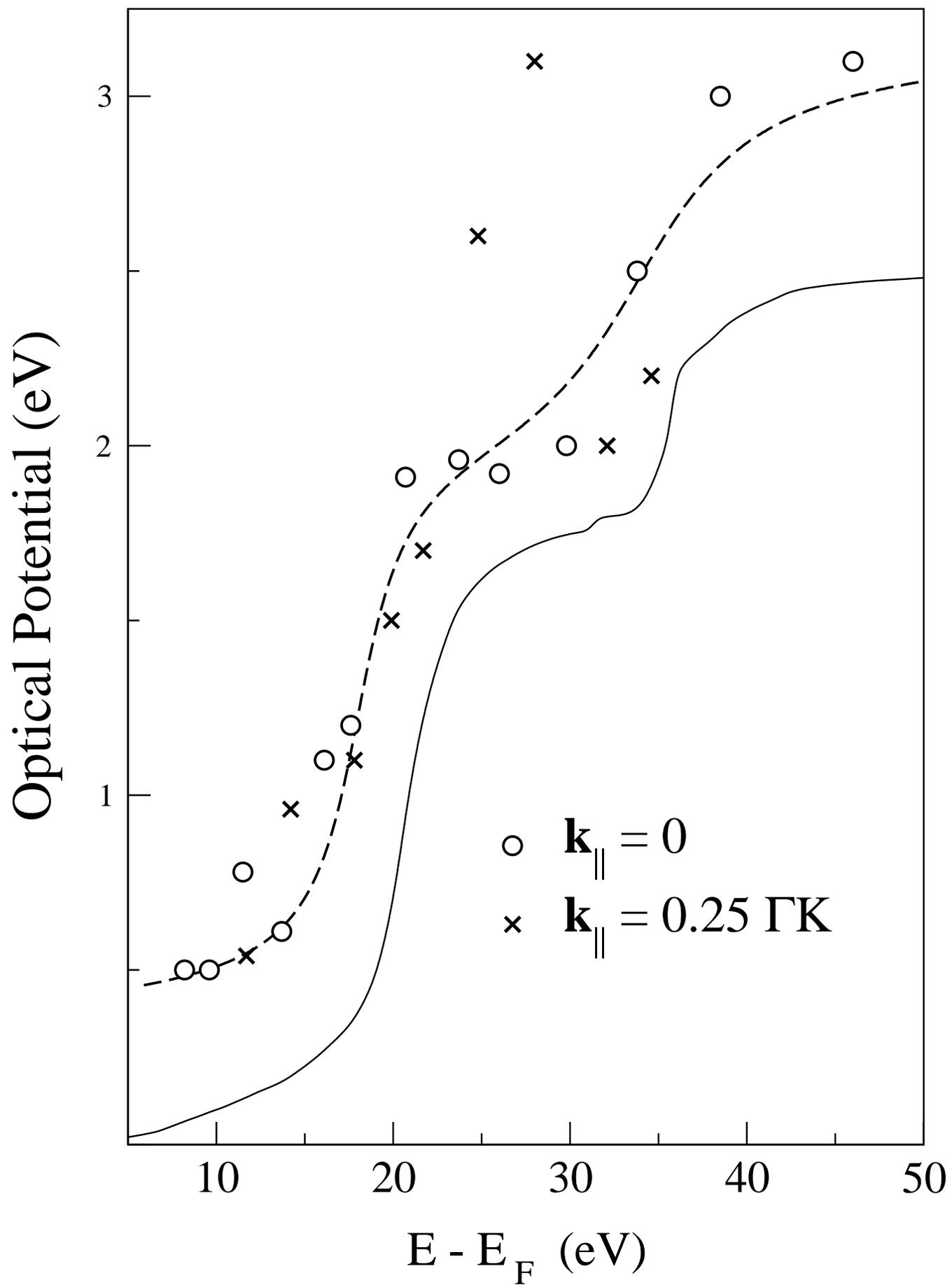
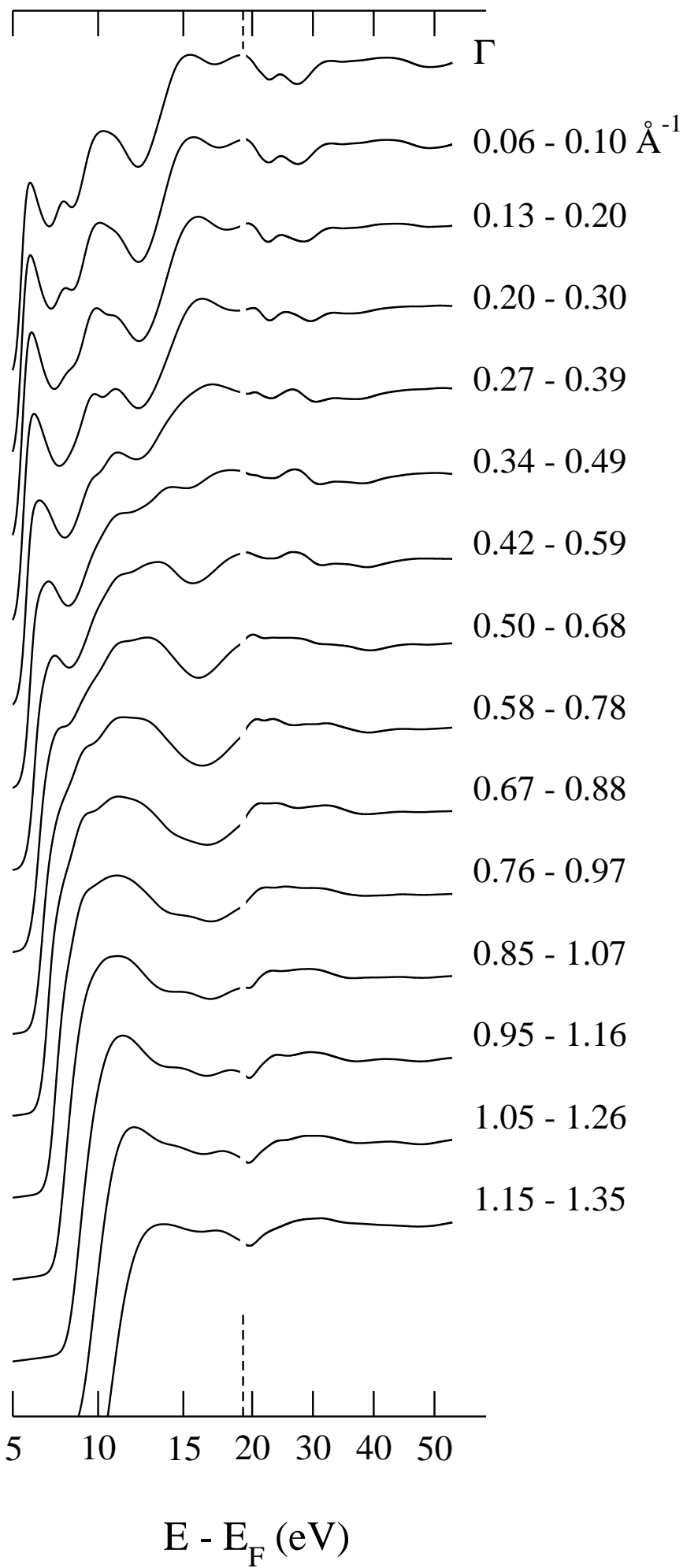
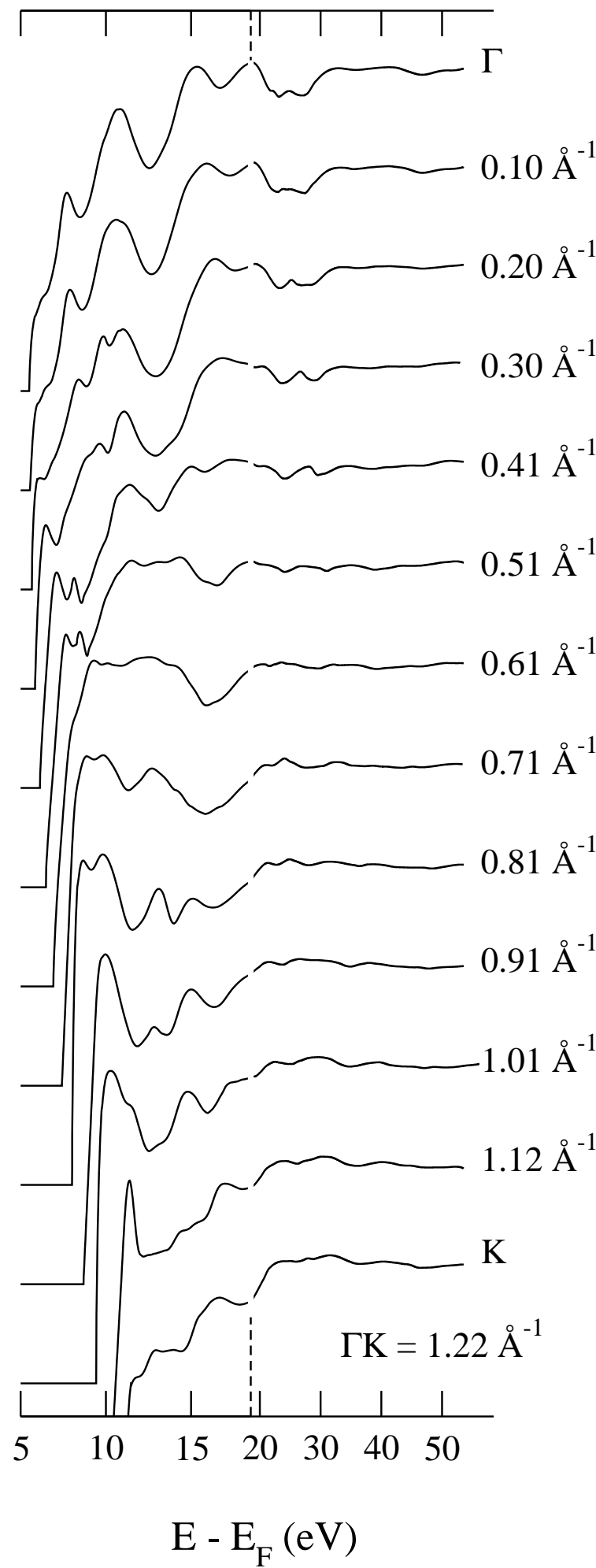
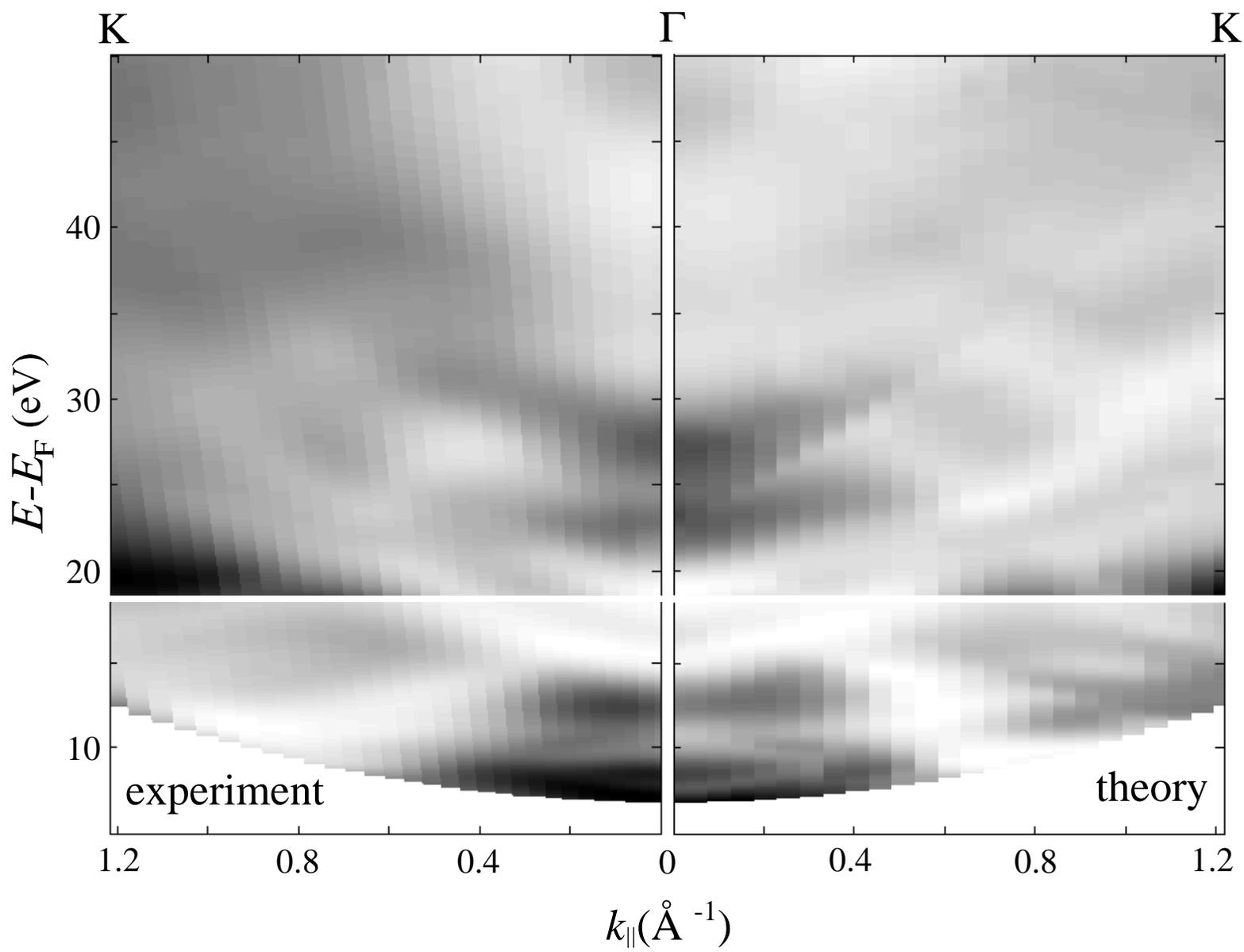
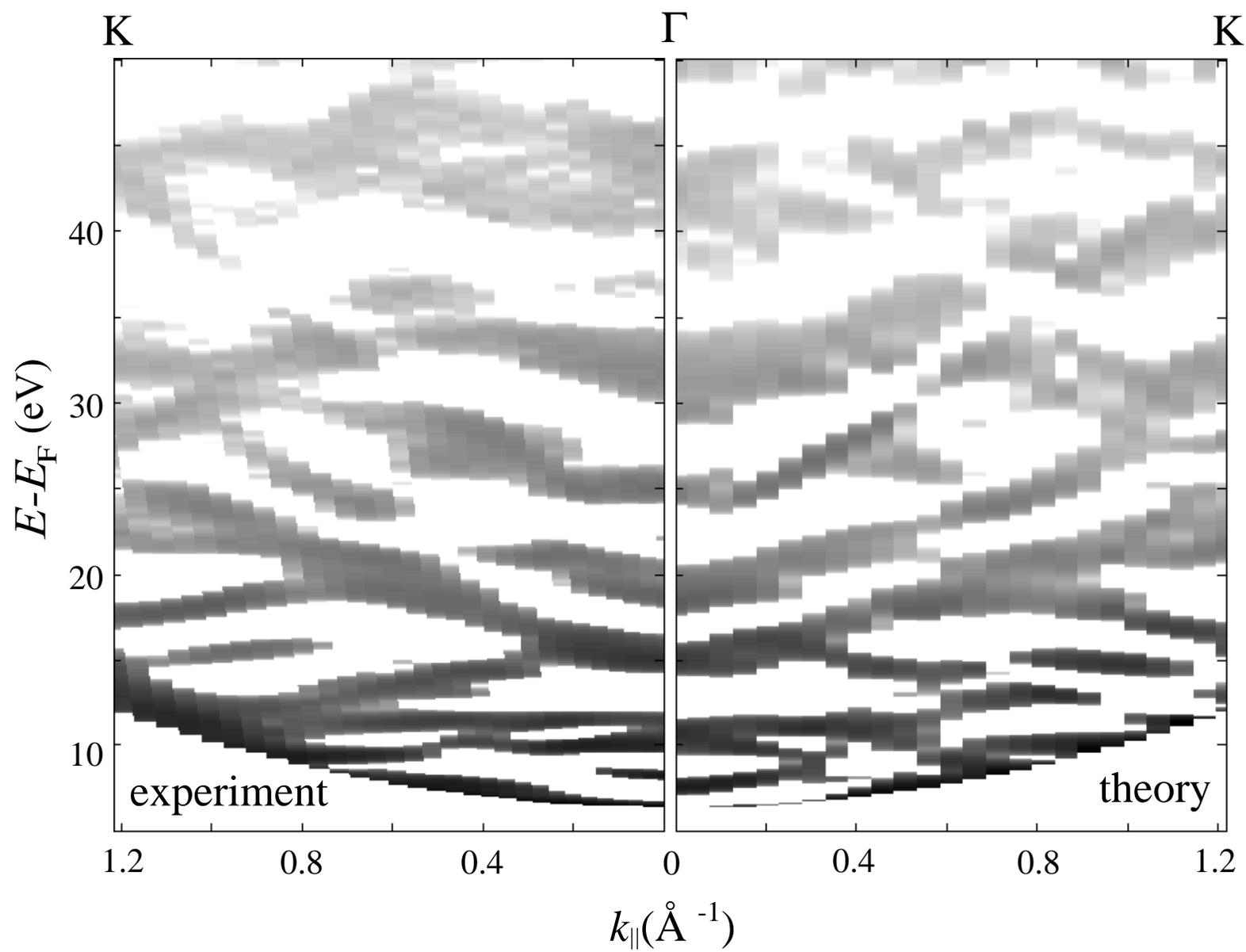
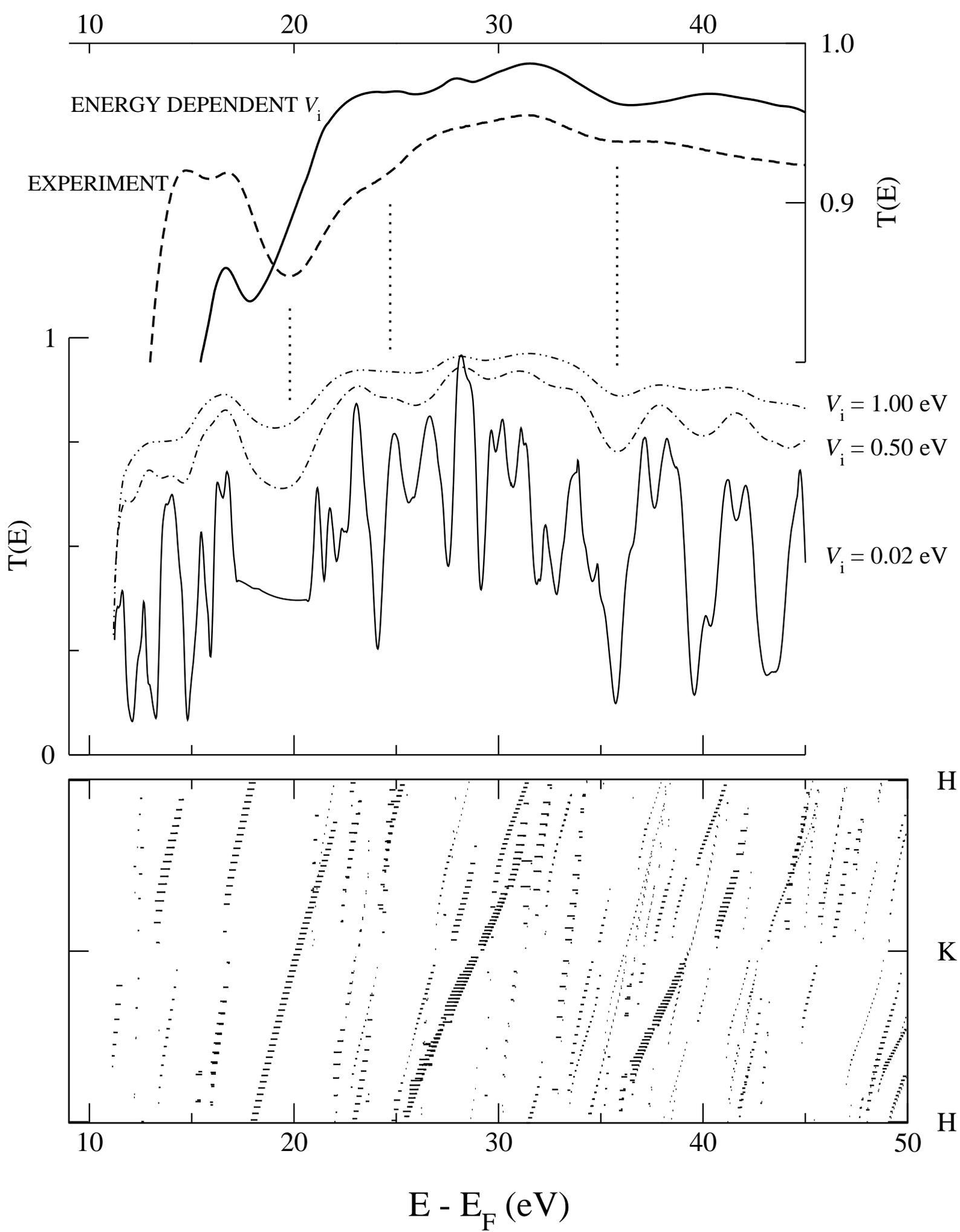
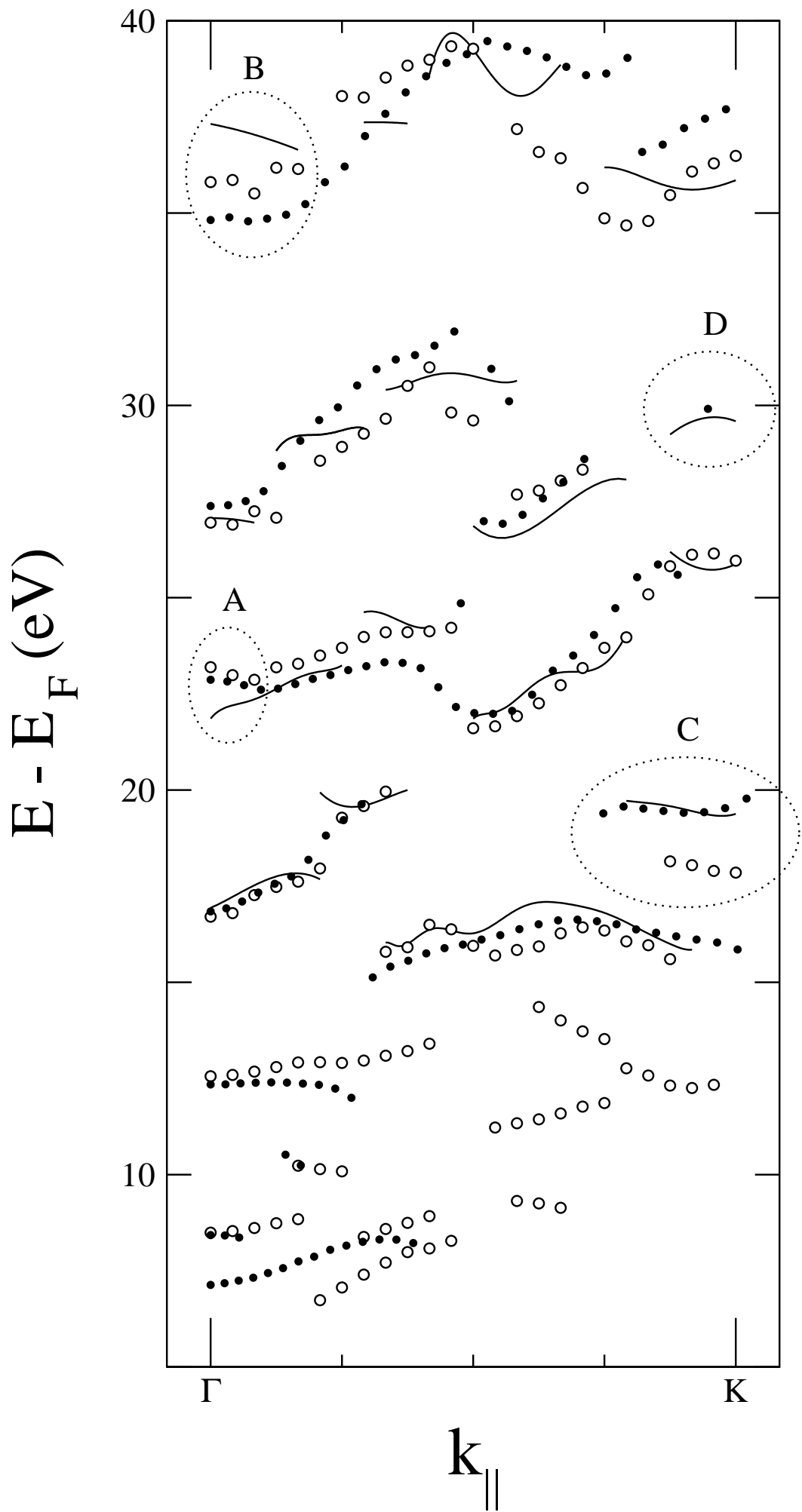
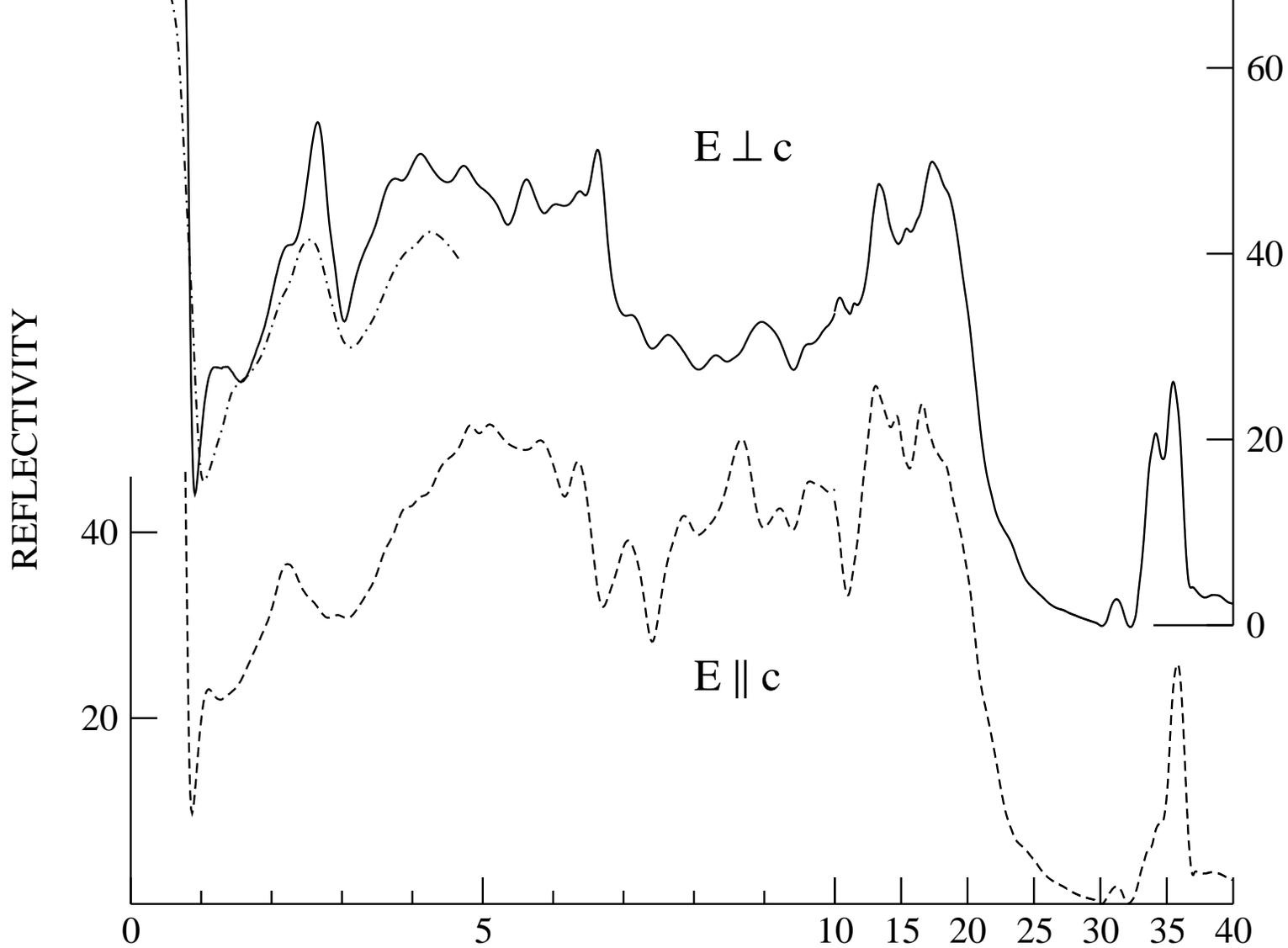
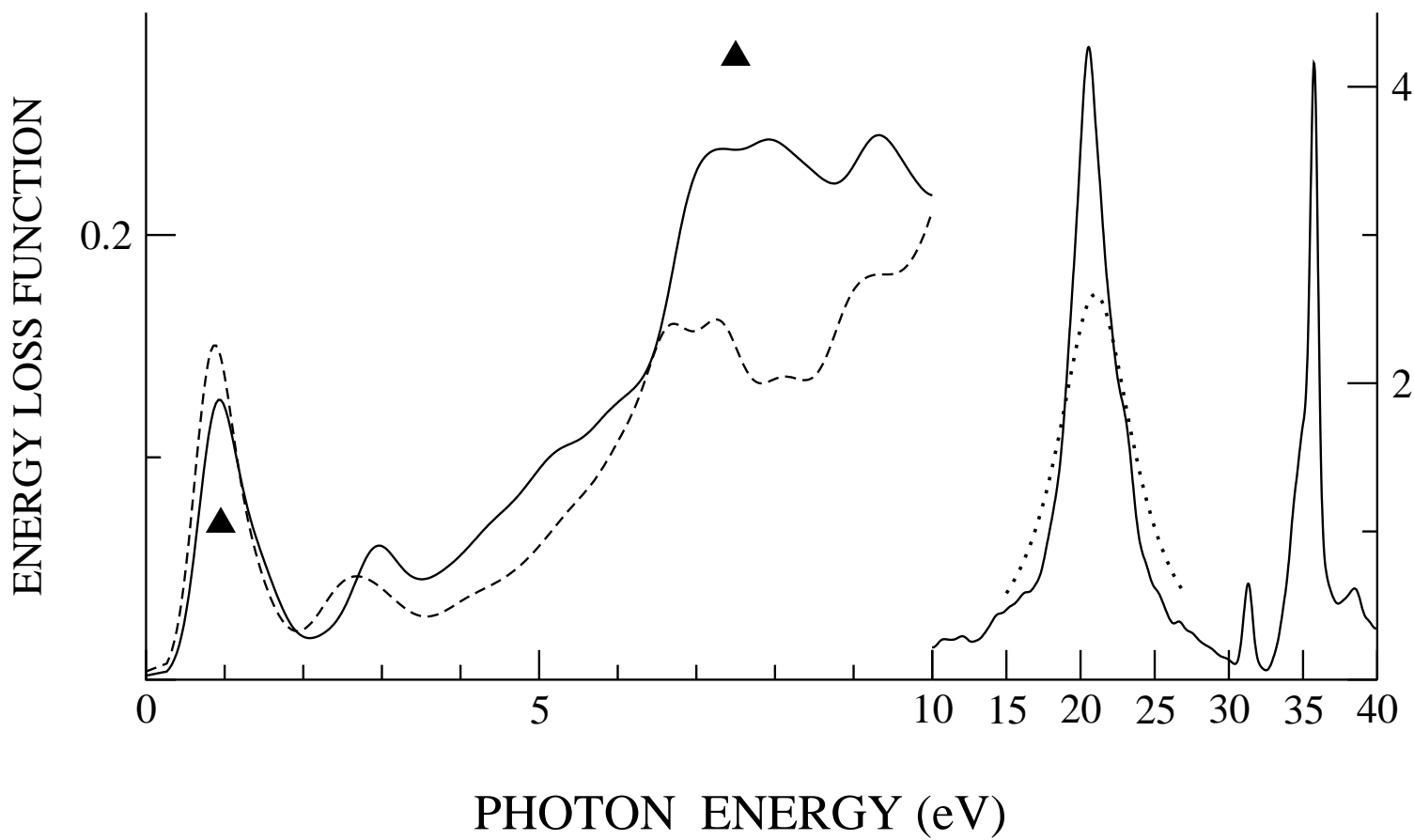
\end{document}